\documentstyle[12pt]{article} 
\begin{document} 
\font\titfont=cmbx10 scaled \magstep2 
\font\autor=cmr9 scaled\magstep1 
\font\autofont=cmbx6 scaled \magstep1 
\def\qed{\vrule height 1.2ex width 1.1ex depth -.1ex} 
\def\litem{\par\noindent 
               \hangindent=\parindent\ltextindent} 
\def\ltextindent#1{\hbox to\hangindent{#1\hss}\ignorespaces} 
\newcommand{\nz}{\hfill\break\noindent} 
\newcommand{\sn}{\smallskip\noindent} 
\newcommand\mn{\medskip\noindent} 
\newcommand{\bn}{\bigskip\noindent} 
\newcommand{\D}{{\cal D}} 
  
\noindent 
\centerline{{\titfont Gauged ${\bf NJL}$ model at strong curvature}}  
  
\vspace{1.0cm} 
  
\mn 
{\autor\centerline{B. Geyer 
        \thanks{E-mail: geyer@ntz.uni-leipzig.d400.de}}} 
{\autor\centerline{Institute of Theoretical Physics and}} 
{\autor\centerline{Center for Theoretical Sciences}} 
{\autor\centerline{Leipzig University}} 
{\autor\centerline{Augustusplatz 10, D-04109 Leipzig, Germany}} 
\sn 
{\autor\centerline{and}} 
{\autor\centerline{S. D. Odintsov 
        \thanks{E-mail: sergei@ecm.ub.es }}} 
{\autor\centerline{Dept. of Physics and Mathematics}} 
{\autor\centerline{Tomsk Pedagogical University}} 
{\autor\centerline{634041 Tomsk, Russia}} 
  
\vspace{1.0cm} 
  
\begin{abstract} 
  
We investigate the gauged NJL--model in curved spacetime using the RG  
formulation and the equivalency with the gauge Higgs--Yukawa model in a 
 modified $1/N_c$--expansion. The strong curvature induced chiral symmetry 
 breaking is found in the non-perturbative RG approach (presumably equivalent 
 to the ladder Schwinger--Dyson equations). Dynamically generated fermion  
mass is explicitly calculated and inducing of Einstein gravity is briefly  
discussed. This approach shows the way to the non-perturbative study 
of the dynamical symmetry breaking at external fields. 
  
\vspace{0.5cm} 
  
\vspace{0.5cm} 
{\footnotesize 
\noindent 1~~ E-mail: geyer@ntz.uni-leipzig.d400.de \\ 
\noindent 2~~ E-mail: sergei@ecm.ub.es \\\mbox{}~~~ 
                    \\} 
  
\end {abstract} 
  
\newpage 
  
\section{\large\bf Introduction} 
  
The dynamical symmetry breaking in different quantum field theories  
(like QCD or Standard Model) maybe realized even in the absence of Higgs 
 scalars with the help of the Nambu--Jona-Lasinio mechanism \cite{1}.  
Moreover, NJL--like models admit the analytic treatment of composite  
bound states. Such models can be used as effective theories for the  
Standard Model and GUT's. 
  
It would be of interest to study the cosmological applications of the   
NJL--like models. In particular, in the Early Universe the phase  
structure of the NJL--model will be qualitatively changed due to non--zero  
curvature effects. The study of the non--gauged NJL-- model in curved  
spacetime \cite{2} has shown a quite rich phase structure and the  
possibility of curvature--induced transitions between chiral symmetric 
 and chiral non--symmetric phases. Moreover, the NJL--model  
(in $1/N_c$ -- expansion) maybe useful for the solution of the cosmological  
constant problem via an effective theory for the conformal factor \cite{3}. 
  
The investigation of the more realistic gauged NJL--model is much more  
complicated. Recently, such a model in flat spacetime has been studied  
in full detail in Ref. \cite{4} using the ladder Schwinger--Dyson (SD)  
equations where the renormalizability of the gauged NJL--model also has 
 been discussed. Unfortunately, it is not clear at all how to formulate  
SD equations in curved spacetime \cite{5}. 
  
In Ref. \cite{6} the flat spacetime gauged NJL--model has been discussed  
using the renormalization group (RG) formalism with appropriate compositeness 
 conditions for the running couplings. In Ref. \cite{7} such formulation has  
been used to show that some class of gauge Higgs--Yukawa models in leading  
order with respect to a modified $1/N_c$--expansion leads to a well defined  
non--trivial theory which is equivalent to the gauged NJL--model. In this  
sense, the gauged NJL--model maybe considered as a renormalizable theory and 
 the RG--improved effective potential \cite{8} maybe found \cite{7} in exact 
 agreement with the ladder--SD effective potential \cite{4} (for earlier  
discussion of the ladder SD potential see \cite{9}). 
  
In the present letter, using the approach of Refs. \cite{6}, \cite{7}  
we study the gauged NJL--model in curved spacetime. The RG--improved  
effective potential is calculated at strong curvature and at weak curvature. 
 The possible chiral symmetry breaking in curved spacetime is shown, and the 
 induced Einstein gravity within the gauged NJL--model is described.

\section{\large\bf Renormalization group and gauged NJL--model in 
 curved spacetime} 
  
Let us start from the $SU(N_c)$ gauge theory with scalars and spinors  
in curved spacetime:

\begin{eqnarray} 
L_m =&-&{\frac{1}{4}}~ G^a_{\mu\nu} G^{a\mu \nu} ~+~ {\frac{1}{2}}~ 
g^{\mu\nu}\partial_{\mu}\sigma \partial_{\nu}\sigma 
 ~-~\frac{1}{2}~ m^2\sigma^2~-~{\frac{\lambda}{4}}~\sigma^4 \nonumber\\ 
&-&{\frac{1}{2}}\xi R\sigma^2~+~{\sum\limits^{N_f}_{i=1}}~ \bar{\psi}_i ~i\hat{\D}~ \psi_i ~-~  
{\sum\limits^{n_f}_{i=1}}~y\sigma \bar{\psi}_i\psi_i~.~ 
\end{eqnarray}

\noindent  
where $\sigma$ is a single scalar, $N_f$ fermions $\psi_i$ belong to  
the representation $R$ of $SU(N_c)$, and $y$ is the Yukawa coupling constant. 
 Note that in order for the theory to be multiplicatively renormalizable in  
curved spacetime one has to add to the Lagrangian (1) the Lagrangian of the 
 external gravitational field \cite{5}. However, this Lagrangian will not 
 be relevant in our discussion. 
  
Let us describe now the modified $1/N$--approximation of Ref. \cite{7} for 
 studying the theory (1) at high energies:\\ 
a) The gauge coupling constant is assumed to be small: 
   
\begin{equation} 
{\frac{g^2N_c}{4\pi}}<< 1 ~, 
\end{equation} 
  
\noindent  
and one has to work only in the first non--trivial 
 order on $g^2$.\\ 
b) The number of fermions should be large enough: 
  
\begin{equation} 
N_f\sim N_c~; 
\end{equation} 
  
\noindent  
however, only $n_f$ fermions ($n_f \ll N_f$) have large Yukawa couplings.\\ 
c) The leading order of $1/N_c$--expansion is also considered; this means, 
in particular, that scalar loop contributions should be negligible 
  
\begin{equation} 
|{\frac{\lambda}{y^2}}| \le N_c~ 
\end{equation} 
  
\noindent (for more details, see \cite{7}). 
  
Within the above approximation, the standard one--loop RG equations for the 
 coupling constants are (see for example \cite{10}, \cite{11} for flat  
spacetime and, for coupling constant $\xi$, see \cite{5}):   
  
\begin{eqnarray} 
&{}&{\frac{dg(t)}{dt}}=-~{\frac{b}{(4\pi)^2}}~g^3(t)~, 
\nonumber\\ 
&{ }&{\frac{dy(t)}{dt}}= 
        {\frac{y(t)}{(4\pi)^2}}~[a~y^2(t)-c~g^2(t)]~, 
\nonumber\\ 
&{ }&{\frac{d\lambda (t)}{dt}} =   
        {\frac{u~y^2(t)}{(4\pi)^2}}~[\lambda (t)-y^2 (t)]~, 
\nonumber\\ 
&{ }&{\frac{d\xi(t)}{dt}} =  
        {\frac{1}{(4\pi)^2}}~2ay^2(t)~ 
        \left(\xi(t)-{\frac{1}{6}}\right)~, 
\end{eqnarray} 
  
\noindent where $ b = (11 N_c-4T(R)N_f)/3, c = 6~C_2(R), a = u/4 = 2~n_fN_c$. 
 Here $t = \ln (\mu/\mu_0)$, and for the fundamental representation we have  
$T(R) = 1/2, C_2(R) = (N_c^2-1)/(2N_c)$.   
  
The one--loop effective potential up to linear curvature terms in the 
 above approximation maybe found as follows ($\sigma^2 \gg |R|$; for flat  
spacetime, see \cite{7}, and for curved spacetime, see \cite{12}): 
  
\begin{eqnarray} 
V&=&{\frac{1}{2}} m^2 \sigma^2 + {\frac{\lambda}{4}} \sigma^4 + 
{\frac{1}{2}} 
\xi R \sigma^2 - {\frac{a M^4_F}{2(4\pi)^2}} \left[\ln  
{\frac{M^2_F}{\mu^2}}-{\frac{3}{2}}\right]\nonumber\\ 
&{}& -~{\frac{a R M^2_F}{12(4\pi)^2}} \left[\ln 
{\frac{M^2_F}{\mu^2}}-1\right], 
\end{eqnarray} 
  
\noindent  
where $M_F = y \sigma$ plays the role of the effective fermion mass for 
 the potential (6). 
  
Similarly, in the case when the curvature is strong $|R| \gg \sigma^2$  
(but $|R| > R^2, R_{\mu\nu}^2$--terms), the $\sigma$--dependent part of 
 the effective potential is 
  
\begin{eqnarray} 
V&=&{\frac{1}{2}} m^2 \sigma^2 + {\frac{\lambda}{4}} \sigma^4 + 
{\frac{1}{2}} 
\xi R \sigma^2 - {\frac{a M^4_F}{2(4\pi)^2}} \left[\ln  
{\frac{-R/4}{\mu^2}}-{\frac{3}{2}}\right]\nonumber\\ 
&{}& -~{\frac{a R M^2_F}{12(4\pi)^2}} \left[ \ln 
{\frac{-R/4}{\mu^2}}-1\right], 
\end{eqnarray} 
  
\noindent 
where $R$ is supposed to be negative. The expressions (6), (7) will be  
used below. 
  
The solution of the RG--equation (5) for the gauge coupling is  
($\alpha = g^2/(4\pi)$) 
  
\begin{equation} 
\eta (t)\equiv {\frac{g^2(t)}{g^2_0}}\equiv {\frac{\alpha 
(t)}{\alpha_0}}= 
(1+{\frac{b~ \alpha_0}{2\pi}}t)^{-1}~. 
\end{equation} 
  
\noindent 
It will not be changed in the gauged NJL--model. 
  
The analysis of solutions for Yukawa and scalar couplings in the above 
 approximation has been done in Ref. \cite{7} in full detail. It has been  
shown that the Yukawa coupling must be asymptotically free in order for the  
theory to be non--trivial. The condition of non--triviality and stability leads 
to non--trivial solution for the scalar coupling (actually the solutions lie 
 on the line between the Gaussian fixed point and the fixed point of  
Ref. \cite{13}, where the reduction of couplings takes place \cite{14}, 
 \cite{15}). 
  
Now we will turn to the $SU(N_c)$ gauged NJL--model with four--fermion 
 coupling constant $G$ in curved spacetime 
  
\begin{equation} 
L=-{\frac{1}{4}} G^2_{\mu \nu} + {\sum\limits^{N_f}_{i=1}} 
\bar{\psi} i \hat{\D} 
\psi_i + G ~{\sum\limits^{n_f}_{i=1}} (\bar{\psi}_i \psi_i)^2~. 
\end{equation} 
  
\noindent 
The standard way to study such a model is to introduce an auxiliary 
 field $\sigma$, in order to identify the NJL--model with the Higgs--Yukawa  
model. As it has been shown in \cite{6}, using the RG approach on can put a  
set of boundary conditions for the effective couplings of the gauge  
Higgs--Yukawa model at $t_{\Lambda} = \ln(\Lambda/\mu_0)$ (where $\Lambda$  
is the UV--cut-off of the gauged NJL--model) in order to prove the  
equivalency of the gauged NJL--model with  the gauge Higgs--Yukawa 
model (at all $\Lambda$, and even for $\Lambda \rightarrow \infty$,  
showing the renormalizability of the gauged NJL--model in this sense  
\cite{7}, see also \cite{4}). 
  
Taking into account the explicit expressions for the compositeness  
conditions \cite{6} one can show that the running Yukawa and scalar 
 couplings in the gauged NJL--model are \cite{7} ($g(t)$ is not changing) 
  
\begin{eqnarray} 
&{}& y^2 (t) ={\frac{c-b}{a}}~ g^2 (t)~ \left[ 
1-\left({\frac{\alpha (t)} 
{\alpha (t_\Lambda)}}\right)^{1-c/b} \right]^{-1} 
\equiv  
y^2_\Lambda (t),\nonumber\\ 
&{}& {\frac{\lambda (t)}{y^4(t)}} = 
{\frac{2a}{2c-b}}~~{\frac{1}{g^2 (t)}} 
\left[ 1- \left( {\frac{\alpha (t)}{\alpha (t_\Lambda)}}\right)^ 
{1-2c/b} \right] \equiv {\frac{\lambda_\Lambda 
(t)}{y^4_\Lambda(t)}}~, 
\end{eqnarray} 
  
\noindent 
where $t<t_{\Lambda}, c<b$. For the case of fixed gauge coupling,   
$b \rightarrow + 0$, eqs. (10) are significally simplified \cite{7}. 
  
In addition, one should have also a compositeness condition for the  
mass which is explicitly written in Ref. \cite{7}. For the case  
$b \rightarrow +0$, it will be 
  
\begin{equation} 
m^2 (t) = {\frac{2a}{(4\pi)^2}} \left( 
{\frac{\Lambda^2}{\mu^2}}\right)^w 
y^2_\Lambda (t)~\mu^2 \left[{\frac{1}{g_4(\Lambda)}} - 
{\frac{1}{w}}\right]~, 
\end{equation} 
  
\noindent 
where $g_4(\Lambda)$ is a dimensionless constant defined by 
$G\equiv ((4\pi)^2/a) g_4(\Lambda)/\Lambda^2$, and $w\equiv 
1-\alpha/(2\alpha_c)$, and where $y^2_\Lambda (t)$  is given by (10)  
at $b \rightarrow +0$, and $\alpha_c^{-1} = 3C_2(R)/\pi$.  
The compositeness condition for $\xi(t)$ is given as (see \cite{12};  
for non-gauged NJL-model, see also \cite{16}) 
  
\begin{equation} 
\xi (t)={\frac{1}{6}}~. 
\end{equation} 
  
\noindent 
Thus, we got the description of the gauged NJL--model via running coupling  
constants (using the equivalency with the gauged Higgs--Yukawa model). 
  
\section{\large\bf RG improved potential and dynamical symmetry breaking} 
  
Let us turn now to the question of dynamical symmetry breaking in the model  
under consideration. For this purpose one has to calculate the effective  
potential, using, for example, the ladder SD equation \cite{4}, \cite{9}.  
Unfortunately, as it was already mentioned it is not clear at all, how  
to formulate SD equations in curved spacetime. However, one may use the  
RG technique because we have the RG formulation of the gauged NJL--model  
via equivalent gauge Higgs--Yukawa model. Surprisingly, it has been shown  
\cite{7} that the RG improved effective potential \cite{8} gives the same  
results as the ladder SD effective potential \cite{4}. 
  
The technique to study the RG improved effective potential is quite  
well--known in flat space \cite{8} as well as in curved space \cite{17}.  
Hence, we will not discuss it in detail here. Using the fact that the  
effective potential satisfies the RG equation, one can explicitly solve  
this equation by the method of characteristics as follows: 
  
\begin{equation} 
V(g,y,\lambda,m^2,\xi,\sigma,...,\mu)= 
V(\bar{g}(t), \bar{y}(t),\bar{\lambda}(t),\bar{m}^2(t), 
\bar{\xi}(t),\bar{\sigma}(t),..., \mu e^t), 
\end{equation} 
  
\noindent 
where the effective coupling constants  
$\bar{g}(t), \bar{y}(t),...,\bar{\xi}(t)$ are defined by the RG  
equations (10), (11) and (12) at scale $\mu e^t$, and $\bar{\sigma}(t)$  
is written in \cite{7} ($t$ is left unspecified for the moment).  
Note that in (13) the gravitational coupling constants connected  
with the Lagrangian of the external gravitational field are not  
written explicitly. These coupling constants are not relevant in  
our discussion because we work in linear curvature approximation.  
Notice also that as boundary condition in (13) it is convenient to  
use the one--loop effective potential (6),(7). 
  
In the following, for simplicity, we restrict ourselves to the fixed  
gauge coupling case $b \rightarrow +0$. (One can also dicuss the general  
case without any problems; however, then the expressions are getting  
very complicated and also non--explicit ones). Chosing the condition  
of vanishing of logarithmic terms in the effective potential (7) for  
finding $t$ we will get: 
  
\begin{equation} 
e^t=\left(\frac{-R}{4\mu^2}\right)^{1/2}~. 
\end{equation}

Using (10) -- (12), (14) in the RG--improved potential (13) with the  
effective potential (7) as boundary condition, one can evaluate the RG  
improved effective potential at strong curvature as follows (the  
convenient RG invariants are given in \cite{7}, \cite{12}): 
  
\begin{eqnarray} 
{\frac{(4\pi)^2}{2a}}~{\frac{V}{\mu^4}} 
&=&     {\frac{x^2}{2}} 
        \left({\frac{\Lambda^2}{\mu^2}}\right)^w 
        \left[{\frac{1}{g_4(\Lambda)}}-{\frac{1}{w}}\right]~ 
\nonumber\\ 
&{}&+~{\frac{x^4}{4}} 
  \left({\frac{-R}{4\mu^2}}\right)^{\frac{-\alpha}{\alpha_c}} ~\left[{\frac{3}{2}} 
  +     {\frac{\alpha_c}{\alpha}} 
  -  {\frac{\alpha_c}{\alpha}} 
                \left({\frac{\sqrt{-R}}{2\Lambda}} 
                \right)^{\frac{2\alpha}{\alpha_c}}   
  \right]~ 
\nonumber\\ 
&{}&+~{\frac{R x^2}{12\mu^2}} 
        \left({\frac{-R}{4\mu^2}} 
        \right)^{{\frac{-\alpha}{2\alpha_c}}}~ 
        \left[{\frac{1}{2}} + {\frac{\alpha_c}{\alpha}}  - {\frac{\alpha_c}{\alpha}} 
          \left({\frac{\sqrt{-R}}{2\Lambda}}\right)^{\frac{\alpha}{\alpha_c}}  
         \right]~, 
\end{eqnarray} 
  
\noindent 
where $x=y_\Lambda(\mu)\sigma_\Lambda(\mu)/\mu$. The expression (15)  
gives the RG improved effective potential for the gauged NJL--model  
(with finite cut-off) at strong curvature. 
  
Taking the limit $\Lambda \rightarrow \infty$ we obtain for the  
renormalized effective potential (see \cite{4}, \cite{7} for a  
discussion of renormalization of the four--fermion coupling constant): 
  
\begin{eqnarray} 
{\frac{(4\pi)^2}{2a}}~{\frac{V}{\mu^4}} 
&=&     {\frac{x^2_{\ast}}{2}} 
        \left[{\frac{1}{g_{4R}(\mu)}}- 
                {\frac{1}{g_{4R}^{\ast}}}\right]  
+~{\frac{x^4_{\ast}}{4}} 
  \left({\frac{-R}{4\mu^2}}\right)^{\frac{-\alpha}{\alpha_c}} 
 ~\left[ {\frac{3}{2}}+{\frac{\alpha_c}{\alpha}} \right]\nonumber\\ 
&{}& 
+~{\frac{R x^2_{\ast}}{12\mu^2}} 
\left({\frac{-R}{4\mu^2}}\right)^{\frac{-\alpha}{2\alpha_c}} 
        \left[{\frac{1}{2}} + {\frac{\alpha_c}{\alpha}}\right]~,    
\end{eqnarray} 
  
\noindent 
where $x_\ast = y_{\ast}\sigma(\mu)/\mu$ and $R < 0$. 
  
Taking into account that  
$\alpha_c/\alpha \simeq (2\pi/3) (4\pi/N_c g^2) \gg 1$ one can study  
the possible chiral symmetry breaking in the model under discussion.  
>From Eqs. (16), taking the first derivative with respect to $x_\ast$, we will find: 
  
\begin{equation} 
x^2_{\ast} = 
        - {\frac{\alpha}{\alpha_c}} 
                \left[{\frac{1}{g_{4R}(\mu)}}- 
                {\frac{1}{g_{4R}^{\ast}}}\right]  
        \left({\frac{-R}{4\mu^2}} 
        \right)^{{\frac{\alpha}{\alpha_c}}}~ 
        - {\frac{R}{6\mu^2}}             
                \left({\frac{-R}{4\mu^2}} 
                \right)^{{\frac{\alpha}{2\alpha_c}}} 
\end{equation} 
  
\noindent 
Hence, even in the case when there is no chiral symmetry breaking in  
flat space (i.e. $g^{-1}_{4R}(\mu) - (g^{\ast}_{4R}(\mu))^{-1} \geq 0$)  
one may expect that chiral symmetry breaking occurs at strong curvature  
(if the second term in (17) is the dominant one). One can also find the  
critical curvature between symmetric and non-symmetric phases. Hence, we  
found the condition defining the chiral symmetry breaking at strong  
curvature (the right--hand side of eq. (17) should be positive). Taking  
into account the fact that in flat spacetime the dynamical chiral symmetry  
breaking in ladder SD approach \cite{4} maybe re--obtained using the  
RG--improved effective potential \cite{7} we are led to the following  
conjecture: Eqs. (16), (17) on dynamical symmetry breaking in the  
NJL--model in curved spacetime should be the same as in the ladder SD  
formulation (which is not developed in curved spacetime yet). It is  
quite remarkable that non--perturbative results on dynamical symmetry  
breaking maybe extended to curved spacetime. Presumably our approach  
maybe useful also for quantum gravity in frames of $1/N$--expansion. 
  
Finally, on the same way as above one can find the renormalized  
effective potential in the situation when $\sigma^2 \gg |R|$ (i.e. using  
eq. (6) as boundary condition). Explicitly, for $\Lambda \rightarrow \infty$  
we have (see also \cite{12}): 
  
\begin{equation} 
{\frac{(4\pi)^2 V}{2a\mu^4}}= 
        {\frac{x^2_{\ast}}{2}} 
        \left[{\frac{1}{g_{4R}(\mu)}}- 
                {\frac{1}{g_{4R}^{\ast}}}\right]  
+{\frac{x^{4/(2-w)}_{\ast}}{4}} 
   \left[ {\frac{3}{2}}+{\frac{\alpha_c}{\alpha}} \right] 
+{\frac{R x^{2/(2-w)}_{\ast}}{12\mu^2}} 
        \left[{\frac{1}{2}} + {\frac{\alpha_c}{\alpha}}\right].     
\end{equation} 
  
\noindent 
This potential again leads to the dynamical symmetry breaking in  
weakly--curved spacetime. The dynamical fermionic mass is given in  
this case as follows: 
  
\begin{equation} 
      x^2_{\ast} \simeq 
        - {\frac{\alpha}{\alpha_c}} 
                \left[{\frac{1}{g_{4R}(\mu)}}- 
                {\frac{1}{g_{4R}^{\ast}}}\right]  
        - {\frac{R}{6\mu^2}}.            
\end{equation} 
  
\noindent 
Hence, we showed the possibility of dynamical symmetry breaking at  
strong curvature as well as at weak curvature. 
  
\section{\large\bf Induced Einstein gravity} 
  
It is quite evident that the NJL--model potential under discussion  
supports the inflationary universe in the same way as the usual Higgs  
scalar potential. Indeed, we have the flat effective potential of $\sigma$  
which gives the effective cosmological constant by its minimum; but such  
vacuum energy serves as a source of inflation. 
  
Here, we will discuss the gravitational effective action taking into  
account $V(\sigma)$ since such an action maybe useful in the study of  
inflation based on induced gravity (due to NJL--model). At GUT's epoch,  
the linear curvature approach is enough to study the applications of the  
effective potential to quantum cosmology. Hence, the gravitational  
effective action in gauged NJL-model (for fixed gauge coupling) maybe  
taken as follows

\begin{equation} 
S = \int d^4x \sqrt{-g} 
\left[\Lambda - {\frac{1}{\kappa^2}}R - V(\sigma)\right] 
\end{equation} 
  
\noindent 
where $V(\sigma)$ is given by eq. (18), and $\Lambda, \kappa^2$ denote the 
 cosmological and gravitational constants due to other effects. 
  
>From (20) one can obtain the induced cosmological and gravitational  
coupling constants as follows: 
  
\begin{eqnarray} 
&{ }& \Lambda_{ind}  
\simeq \Lambda 
        -{\frac{a\mu^4}{(4\pi)^2}}       
         \left[\left({\frac{1}{g_{4R}(\mu)}}- 
                        {\frac{1}{g_{4R}^{\ast}}}\right) x^2 
                        + {\frac{\alpha_c}{2\alpha}} x^4\ 
        \right]~, 
\nonumber\\ 
&{ }& {\frac{1}{\kappa^2_{ind}}}  
\simeq{\frac{1}{\kappa^2}}  
        + {\frac{a\mu^2}{6(4\pi)^2}}{\frac{\alpha_c}{\alpha}} x^2. 
\end{eqnarray} 
  
\noindent 
Hence, Einstein gravity maybe induced by purely quantum effects in the  
gauged NJL--model (i.e. even for $\Lambda = \kappa^{-2} = 0$). In this  
case, $\Lambda_{ind}, \kappa^{-2}_{ind}$ maybe easily defined by the  
minimum of the effective potential using (19). 
  
Thus, induced Einstein gravity maybe succesfully obtained within the  
gauged NJL--model (for induced Einstein gravity in GUT's see, for example,  
\cite{5}). It would be of interest to investigate the details of  
inflationary scenario in such a model. 
  
In summary, using the RG language and the equivalency with the gauge  
Higgs--Yukawa model in a modified $1/N_c$--expansion we studied the  
gauged NJL--model in curved spacetime. Applying the RG improved  
effective potential the curvature--induced chiral symmetry breaking  
is investigated and (presumably non--perturbative) dynamically  
generated fermionic mass is found at strong and at weak curvature.  
The possiblity of inducing Einstein gravity is briefly discussed. 
  
It should be of interest to discuss the same model at some specific  
background (like DeSitter universe) what maybe useful for developing 
 of a "four--fermion model of inflation". 
  
>From another side the above approach maybe useful to study the  
non--perturbative aspects of quantum gravity (in particular,  
$R^2$--gravity \cite{5}) interacting with the gauged NJL--model.  
This question deserves further study. 
  
This work has been supported by MEC (Spain), RFBR (Russia) 
project 94-020324, and by "Acciones Integradas Hispano--Alemanas". 
  
\end{document}